\documentclass{epl}

\title{Nonlinear electrodynamics and the Pioneer 10/11 spacecraft
anomaly}

\author{Jean Paul Mbelek$^0$ \and Herman J. Mosquera Cuesta$^{1,2}$
\and M. Novello$^1$ \and Jos\'e M. Salim$^1$}

\institute{\inst{0}Service d'Astrophysique, C. E. Saclay, 91191 Gif-sur-Yvette 
Cedex, France\\
\inst{1}Instituto de Cosmologia, Relatividade e Astrof\'{\i}sica
(ICRA-BR), Centro Brasileiro de Pesquisas F\'{\i}sicas (CBPF)\\
Rua Dr.  Xavier Sigaud 150, 22290-180,  RJ, Brasil :::  hermanjc@cbpf.br \\
\inst{2}{Abdus Salam International Centre for Theoretical Physics, 
Strada Costiera 11, Miramare 34014, Trieste, Italy}
}


\begin{document}

\newcommand*{\be}{\begin{equation}}
\newcommand*{\ee}{\end{equation}}

\maketitle

\begin{abstract}
The occurrence of the phenomenon known as photon acceleration is a natural
prediction of nonlinear electrodynamics (NLED). This would appear as an
anomalous frequency  shift in any modelization of the electromagnetic field
that only takes into account  the classical Maxwell theory. Thus, it is 
tempting to address the unresolved anomalous, steady, but time-dependent,
blueshift of the Pioneer 10/11 spacecrafts within the framework of NLED.
Here we show that astrophysical data on the strength of the magnetic
field in both the Galaxy and the local (super)cluster of galaxies support 
the view on the major Pioneer anomaly as a consequence of the phenomenon 
of photon acceleration. If confirmed, through further observations or lab
experiments, the reality of this phenomenon should prompt to take it into 
account in any forthcoming research on both cosmological evolution and  
origin and dynamical effects of primordial magnetic fields, whose seeds 
are estimated to be very weak. %%% Date: 08/Oct/2006   :::  11:25'
\end{abstract}

%%% newcommand*{\.}{\'{\i}

%%%%   Le Plan du Papier
%%%% --------------------

%%%  Phenomenon ----> Pioneer 10/11
%%%   a) Plusieres de Theories --- > donnant explications le effect
%%%      Pioneer
%%%   Il'ya aucune theorie que utilise le NLED !
%%%   On peux liee Pioneer < --- > NLED (L (F) = -1/2 F + Gamma/F
%%%      a) On a la metrique effective (g^{\mu \nu})_eff =
%%%         L_F G g^{\mu \nu} - L_{FF} F^\mu_\alpha F^{\beta \nu}
%%%         k^\alpha k_\beta
%%%    dot{\nu}{\nu} =  dot{\nu}{\nu}(H_0, B_0/B_n)
%%%    Donc  dot{\nu}{\nu} < ---- . Pioneer effect
%%%    Implications

\section{The Pioneer 10/11 spacecraft anomaly}

Since 1998, Anderson {\sl et al.}  have continuously reported an anomalous
frequency shift derived from about ten years study of radio-metric data from
Pioneer 10: 03/01/1987-22/07/1998, Pioneer 11: 05/01/1987-01/10/1990, and of 
Ulysses and Galileo spacecrafts \cite{1}.  The observed 
effect mimics a constant acceleration acting on the spacecraft with magnitude
$a_{P} = (8.74 \pm 1.33) \times 10^{-8}$~cm~s$^{-2}$ and a steady frequency 
drift $\frac{d \Delta \nu}{dt} \simeq 6 \times 10^{-9}$~Hz/s which equates to 
a "clock acceleration": $\frac{d\Delta \nu}{dt} = \frac{a_{P}}{c}~\nu 
\label{eq 0}$ ($\ddag$), where $t$ is the one way signal travel time. An
independent analysis %% of radio Doppler tracking data from the Pioneer 10 spacecraft 
for the period 1987 - 1994 confirms the previous observations 
\cite{2}. In addition, by removing the spin-rate change contribution yields an 
apparent anomalous acceleration $a_{P} = (7.84 \pm 0.01) \times$ ~$10^{-8}$ cm~
s$^{-2}$, of the same amount for both Pioneer 10/11 \cite{3,8}. Besides, it has 
been noted that the magnitude of $a_{P}$ compares nicely to $cH_{0}$, where 
$H_{0}$ is the Hubble parameter today. 

Unlike other spacecrafts as the Voyagers and Cassini which are three-axis
stabilized (hence, not well-suited for a precise reconstitution of trajectory
because of numerous attitude controls), the Pioneer 10/11, Ulysses and the
by-now destroyed Galileo are attitude-stabilized by spinning about an axis
(parallel to the axis of the high-gain antenna) which permits precise 
acceleration estimations to the level of $10^{-8}$~cm~s$^{-2}$ (single 
measurement accuracy averaged over $5$ days). Besides, because of the 
proximity of Ulysses and Galileo to the Sun, the data from both spacecrafts 
were strongly correlated to the solar radiation pressure unlike the data from 
the remote Pioneer 10/11. Let us point out that the motions of the four 
spacecrafts are modelled by general relativistic equations (see \cite{3}, 
section $IV$) including the perturbations from heavenly bodies as small as 
the large main-belt asteroids (the Sun, the Moon and the nine planets are 
treated as point masses). Proposals for dedicated missions to test the Pioneer 
anomaly are now under consideration \cite{proposals}.
In search for a possible origin of the anomalous blueshift, a number of
gravitational and non-gravitational potential causes have been ruled out 
by Anderson {\sl et al} \cite{3}. According to the authors, none of these 
effects may explain $a_{P}$ and some are $3$ orders of magnitude or more 
too small. The addition of a Yukawa force to the Newtonian does not work 
easily. An additional acceleration is predicted by taking into account 
the Solar quadrupole moment \cite{4}. Although this entails a blueshift, 
it decreases like the inverse of the power four of the heliocentric radius, 
being of the order of $a_{P}$ only below $2.1$~AU.
Meanwhile, the claim that the Modified Newtonian Dynamics (MOND) may
explain $a_{P}$ in the strongly Newtonian limit of MOND \cite{6,7} is not
obvious at all. %%% First, the fits to the rotational curves of spiral galaxies 
%%yield for the MOND acceleration constant $a_{0}$ a value eight times smaller 
%%than $cH_{0}$ \cite{6}. B Second, the gravitational pulling of the Sun up to 
%%$100$ AU still yields an acceleration greater than $a_{0}$ by at least three 
%%orders of magnitude, equating $a_{0}$ only at about $3000$ AU. B Hence, Newtonian
%%dynamics up to general relativity corrections should apply to the spacecrafts.
%%Otherwise, one would be inclined to conclude that MOND is ruled out by a 
%%laboratory experiment \cite{7}.
Therefore, the alternative that the Pioneer anomaly does not result from a
real change in velocity (see \cite{3}, section X) deserves to be investigated.

Indeed, a direct interpretation of the observational data from the spacecrafts 
implies merely an anomalous time-dependent blueshift of the photons of the 
communication signals. On the other hand, in using a time dependent potential \cite{12,13} 
to explain the Pioneer 10/11 data one may be pointing out to the need of an effective 
metric for the photons. In fact, what is needed is just a time variation of the 
$4$-momentum of the photon along its path.  Thus the atomic energy levels would 
not be affected, only the motion of the photon being concerned.

{ 
In summary, prosaic explanations, non-gravitational forces and modified 
dynamics or new interaction (long or short range) force terms do not work\cite{4,6,7,10}.
Gravitational origin of the anomaly is rouled out by the precision of the 
planetary ephemeris (see Anderson et al. \cite{1}, Iorio \cite{iorio2006},  
and others \cite{tangen2006}) and the known bounds on dark matter within 
the orbital radius of Uranus or Neptune \cite{DM}.
Hence, the Pioneer anomaly seems not to be related to the gravitational 
\cite{1,iorio2006,tangen2006}, but rather to the EM sector (since these 
two are the only long range interactions known today). Non-metric fields 
can also be regarded as gravitational fields and there is a lot of space 
for speculation. The possibility of an 
interaction of the EM signal with the solar wind leading to a change of the 
frequency of the EM signal is now rouled out (see Anderson et al. \cite{3}). 
It is clearly the equation of
motion of the photon that is concerned, that is, what happens to the photon 
during its propagation from the Pioneer 10/11 antennas to the receivers 
on Earth. Now, classical (Maxwell theory) or quantized (QED) linear 
electrodynamics does not allow for a change of the frequency of a photon 
during its propagation in a linear medium without invoking diffusion due 
to the interaction with the surrounding matter (hence a smear out of the 
image of the source). Indeed, for such a phenomenon to occur, one needs 
to consider a general Lagrangian density $L = L(F)$ for which its second 
derivative w.r.t. $F$: $d^2L/dF^2 = L_{FF} \neq 0$. Therefore, the Pioneer 
anomaly, if not an artifact, may be a result of NLED as we show below. Indeed, 
relation ($\ddag$) above translates in covariant notation into $\frac{dx^{\nu}}
{d{\sl l}}~\nabla_{\nu}~k^{\mu} = \frac{a_{P}}{c^2}~k^{\mu}$, where ${\sl l}$ is 
some affine parameter along a ray defined by $k^{\mu} = \frac{dx^{\mu}}{d{\sl l}}$ 
(see \cite{Fujii}). The latter equation departs 
from the classical electrodynamics one $\frac{dx^{\nu}}{d{\sl l}}~\nabla_{\nu}
~k^{\mu} = 0$ (see \cite{LL1970}, section 87) and suggests the NLED effect 
dubbed photon acceleration. The concept of photon acceleration, which
follows from the description of photon propagation in NLED, was introduced 
by Ref.\cite{Novello-Salim2002}. We explain next why the anomaly shows 
up in some situations and not others. For experimental tests of NLED and 
further theoretical predictions see \cite{NLED-REFRENCES}.
}
%%%%%%%%%%%%%%%%%%%%%%%%%%%%%%%%%%%%%%%%%%%%%%%%%%%%%%%%%%%%%%%%%%%%%%%%%%%%%%%%%

\section{NLED and A Lagrangian for All Scales: From Cosmology to the Solar 
System} 

Indeed, all these requirements are achieved by considering 
NLED based on a Lagrangian density $L(F)$ that includes terms depending
non-linearly on the invariant $F = F_{\mu\nu}~F^{\mu\nu}, F = 2 (B^2 c^2 
- E^2)$ \cite{Novelloal2000,Novello-Salim2002,plebanski}, instead of the 
usual Lagrangian density $L = - \frac{1}{4} F$ of the classical 
electromagnetism in a vacuum. Hereafter we investigate the effects 
of nonlinearities in the evolution of EM waves, described onwards as 
the surface of discontinuity of the EM field.  Extremizing the Lagrangian
with respect to the potentials $A_{\mu}$ yields the following
field equation \cite{plebanski}:
%Next we show that the "acceleration" of photons
%predicted by NLED may account for the anomalous blueshift indicated by the
%Pioneer 10/11, Ulysses and Galileo spacecrafts.  This will manifest itself
%as a new frequency shift for the EM waves, in addition to the
%Doppler shift (special relativity) and the gravitational and cosmological
%redshift (general relativity).  
\be
\nabla_{\nu} (L_{F}F^{\mu\nu} ) = 0
\label{eq60} ,
\ee
where $\nabla_\nu $ defines the covariant derivative, and $L_F = 
dL/dF$. Besides this, we have the cyclic identity:
\be
\nabla_{\nu}F^{*\mu\nu} = 0 \hskip 0.3 truecm \Leftrightarrow
\hskip 0.3 truecm F_{\mu\nu|\alpha} + F_{\alpha\mu|\nu} +
F_{\nu\alpha|\mu} = 0\, .
\label{eq62}
\ee
%%%%%   The field equation can be written explicitly as:
%%%% \be
%%%%  \nabla_{\nu}(L_{F}F^{\mu\nu})=0
%%%% \label{eq61} \, .
%%%%  \ee
%%%%    (\ref{eq61})
Taking the discontinuities of the field equation yields\cite{hadamard}
\be
L_{F} f^{\, \, \mu}_{\lambda} k^\lambda + 2L_{FF}F^{\alpha\beta}
f_{\alpha\beta} F^{\mu\lambda} k_{\lambda} = 0 \, .
\label{j1}
\ee
The discontinuity of the  Bianchi identity renders:
\be
f_{\alpha\beta}k_{\gamma} + f_{\gamma\alpha}k_{\beta} +
f_{\beta \gamma} k_{\alpha} = 0 .
\label{bianchi-discont}
\ee
To obtain a scalar relation, we contract Eq.(\ref{bianchi-discont})
with $ k^{\gamma}F^{\alpha\beta} \label{eq25}$, resulting
\be
(F^{\alpha\beta}f_{\alpha\beta} g^{\mu\nu} + 2F^{\mu\lambda}
f_{\lambda}^{\, \, \nu})k_{\mu} k_{\nu}= 0 \, .
\label{j2}
\ee
We have two distinct cases: $F^{\alpha\beta} f_{\alpha\beta} = \chi, \, 
{\rm or} \, 0$. If it is zero, such a mode propagates along standard null 
geodesics. When it is $\chi$, we obtain, from Eqs.(\ref{j1}) and 
(\ref{j2}), the propagation equation for the field discontinuities 
\be
\left(g^{\mu\nu} - 4\frac{L_{FF}}{L_{F}} F^{\mu\alpha}
F_{\alpha}^{\, \, \nu}\right) k_{\mu}k_{\nu} = 0 \, .
\label{63}
\ee
Taking the derivative of this expression, we obtain

\be
k^{\nu} \nabla_\nu k_{\alpha} = 4 \left(\frac{L_{FF}}{L_{F}}F^{\mu\beta}
F_{\beta}^{\, \, \nu} ~k_{\mu} k_{\nu}\right)_{|\alpha}.
\label{kuknu}
\ee

Eq.(\ref{kuknu}) shows that the nonlinear Lagrangian introduces a term acting as 
a force accelerating the photon.

\section{Photon acceleration in NLED}

{ If NLED is to play a 
significant role at the macroscopic scale, this should occur at the 
intermediary scales of clusters of galaxies or the interclusters medium, 
wherein most observations show that the magnetic fields are almost uniform (and 
of the same order of magnitude \cite{GALACTIC-B-FIELDS,vogt-ensslin2003}), 
unlike the dipolar magnetic fields of the Sun and planets. However, galaxies 
are gravitationally bound systems, whereas the cosmic expansion is acting at the 
cluster of galaxies scale. Thus, the magnetic field (${\bf B}$) in clusters 
of galaxies (IGMF) depends on the cosmic time. So, the ${\bf B}$ that is relevant 
to this study is that of the local cluster of galaxies \cite{beck2000}. }
(As for the contribution of the CMB radiation see \cite{estimating-gamma}).
%%% The interclusters magnetic field is in any case by far small ($10^{-9}$
%%% ~G) to add a measurable correction even to the cosmological redshift. 
%As for the contribution of the cosmic microwave background (CMB), not only 
%it is too weak but also, the CMB is pure radiation ($F = 0$), whereas we are 
%interested in the case of a background magnetic field with no significant electric 
%field counter-part, i.e., $E=0$. 
Recently, Vall\'ee \cite{Vallee} has speculated 
that the $2~\mu$G magnetic field he has observed within the local supercluster 
of galaxies in cells of sizes of about $100$~kpc may extend all the way to the 
Sun. We explore further this idea in the framework of NLED and show that it is 
capable to provide an explanation of the Pioneer anomaly from first principles.

Relation (\ref{63}) may be casted in the form 

\be
g_{\mu\nu} k^{\mu}k^{\nu} = 4\frac{L_{FF}}{L_{F}} b^2,
\label{64}
\ee

where $b^\mu = F^{\mu\nu} k_\nu$ and $b^2 = b^\mu b_\mu$. As $E=0$, one can write, 
after averaging over the angular-dependence \cite{BB1970}: $b^2 = - \frac{1}{2}
||\vec{k}||^2 B^2 c^2 = - \frac{1}{4}||\vec{k}||^2 F \label{65}$, with $||\vec{k}||
= \omega/c = 2\pi \nu/c$. By inserting this relation in (\ref{64}) yields

\be
g_{\mu\nu} k^{\mu}k^{\nu} = - \frac{\omega^2}{c^2} \,F \, \frac{L_{FF}}{L_{F}}.
\label{66}
\ee

Taking the $x^{\alpha}$ derivative of Eq.(\ref{66}) we obtain

\be
2 g_{\mu\nu} k^{\mu} (k^{\nu})_{|\alpha} \,+ \,k^{\mu} k^{\nu}
(g_{\mu\nu})_{|\alpha} = - (\frac{\omega^2}{c^2} \, F \, 
\frac{L_{FF}}{L_{F}})_{|\alpha}.
\label{67}
\ee

The cosmological expansion will be represented by $g_{\mu\nu} = a^{2}(\eta) 
g_{\mu\nu}^{(local)} \; \label{68}$, with $a$ the scale factor, $\eta$ the
conformal time, and $g_{\mu\nu}^{(local)}$ the local metric. So, Eq.(\ref{67}) 
yields: $2 g_{\mu\nu} k^{\mu} (k^{\nu})_{|0} \,+ \,2 \frac{\dot{a}}{a} \,g_{\mu\nu}
k^{\mu} k^{\nu} = -\,(\frac{\omega^2}{c^2} \,F \, \frac{L_{FF}}{L_{F}})_{|0}
\; \label{69}(\star)$, where the dot stands for partial derivative w.r.t. $\eta$. Using 
Eqs.(\ref{66}) and ($\star$) we obtain\footnote{By removing the NLED extraterm from 
Eq.(\ref{64}), this reduces it to $g_{\mu\nu}^{(local)} k^{\mu}k^{\nu} = 0$ so that 
the photons would just see the local background metric.}

\be
k_{\mu} (k^{\mu})_{|0} = \frac{\dot{a}}{a} \, \frac{\omega^2}{c^2} \,F
\, \frac{L_{FF}}{L_{F}} \,- \, \frac{1}{2} \, (\frac{\omega^2}{c^2} \,F
\, \frac{L_{FF}}{L_{F}})_{|0}.
\label{70}
\ee

Now, $\dot{F} = \,- \,4 \, \frac{\dot{a}}{a} \,F, \label{71}$ by recalling that 
$B^2 \propto a^{-4}$. Moreover, from the method of the effective metric, it can 
be shown that $k^{0}$ does not vary with time in the first order approximation 
\cite{approximation} unlike $|| \vec{k} ||$. Hence $k_{\mu} (k^{\mu})_{|0} = - \,
\frac{\omega}{c} \,\frac{\dot{\omega}}{c}\; \label{72}(\star\star)$. By inserting 
relation ($\star\star$) in (\ref{70}), and then expanding and arranging, one finds

% Hence $k_{\mu} (k^{\mu})_{|0} = - \, \frac{\omega}{c} \,\frac{\dot{\omega}}{c} \label{72}$. 
%By inserting (\ref{72}) in (\ref{70}), and then expanding and arranging, one finds

\be
\frac{\dot{\nu}}{\nu} = - \, \frac{\dot{a}}{a} \,\frac{Q + 2F Q_{F}}{1 - Q}.
\label{73}
\ee

where we have set $Q = F \, \frac{L_{FF}}{L_{F}}$ and $Q_{F} = \partial
Q/\partial F$.

At present cosmological time ($t$) and for a duration very short as compared 
to the universe age, Eq.(\ref{73}) reduces to $\frac{\dot{\nu}}{\nu} \simeq - \, 
H_{0} \, \frac{Q + 2F Q_{F}}{1 - Q} \label{74}$ ($\dot{\nu}$ is the 
photon frequency $t$-derivative). $\dot{\nu} \neq 0$ if and only if a) the NLED 
contribution is non-null, i.e., $L_{\rm FF} \neq 0 $, and b) $F$ depends on time.

%%%% B  \vskip 0.5 truecm

%%%%%%%%%%%%%%%%%%%%%%%%%%%%%%%%%%%%%%%%%%%%%%%%%%%%%%%%%%%%%%%%%%%%%%%%%%%%%%%%%%%%%%5
%%%%   \vskip 0.5 truecm

\section{The NLED Lagrangian}

The explicit form of this general nonlinear Lagrangian (which simulates the 
effect of dark energy in Ref.\cite{NSB2004}) reads

\be
L = - \frac{1}{4} F + \frac{\gamma}{F} \,  , \hskip 0.3
truecm {\rm or  } \hskip 0.3 truecm L = - \frac{1}{4} F +
\frac{\gamma_n}{F^n} \, ,
\label{12}
\ee

where $n$ is a strictly positive integer. From 
Eqs.(\ref{73},\ref{12}), the time variation of the photon frequency, due to 
interaction with very weak {\bf B}$(t)$ fields, reads

%%%% \begin{widetext}
\begin{equation}
\frac{\dot{\nu}}{\nu} =  A_n \gamma_n \frac{4n\gamma_n - (2n+1)
F^{n+1} }{(F^{n+1} + 4 n \gamma_n) (F^{n+1} + 4n(n+2) \gamma_n)}.
\label{14}
\end{equation}
%%%% \end{widetext}

with $A_n = 4H_0 n(n+1)$. Notice that $\gamma_n$ should be negative in order
to guarantee that the Lagrangian is bound from below (see \cite{LL1970},
sections 27 and 93), $\gamma_n = - (B_n c)^{2(n+1)}$. Also, it is worth 
noticing that Eq.(\ref{14}) in the nearly-zero field limit ($B \rightarrow 0$) 
would reduce to

% which involves $Q > 0$. Since relation (\ref{66}) implies $Q < 1$ and the theory 
%% must be such that $L_{\rm F} < 0$ (hence, $F^{n+1} + 4n \gamma_n > 0$) for the 
%% energy density of the EM field be positive definite (see \cite{Novelloal2000},
%% appendix B), one can verify that the equation (\ref{14}) implies a blueshift.

%%%%%%%%%%%%%%%%%%%%%%%%%%%%%%%%%%%%%%%%%%%%%%%%%%%%%%%%%%%%%%%%%%%%%%%%%%%%

%%%%  \begin{widetext}
\begin{equation}
\frac{\dot{\nu}}{\nu} =  H_0 \frac{n+1}{n+2}\, , 
\label{16}
\end{equation}
%%% \end{widetext}

which implies a blueshift.

\section{Discussion and conclusion}

{ We stress that the NLED is a universal theory for the electromagnetic field, with 
$\gamma_{n=1} = \gamma$ in Eq.(\ref{12}) being a universal constant, the value of
which was fixed by Ref.\cite{NSB2004} by using the CMB constraint. 
%% the observational fact concerning the transition from decelerate-to-accelerate 
%  universe. 
Setting $B_1 = \frac{1}{c} |\gamma|^{1/4}$, one finds $B_1 = 0.008 \pm 0.002 
~\mu$G {\cite{estimating-gamma}}. But be awared of that a conclusive fashion of 
fixing $\gamma$ should benefit of a dedicated laboratory experiment, as it was 
done, for instance, to fix the electron charge through Millikan's experiment. 

Thence, to compute the effect (shift) on the Pioneer communication signal 
frequencies (uplink and downlink), we need only to introduce the value of the 
strength of the local supercluster {\bf B}-field: $B_{\rm LSC} \sim 10^{-8} - 
10^{-7}$~G \cite{Blasi1999}. Now, the theory must be such that $L_{\rm F} < 0$ 
(hence, $F^{n+1} + 4n \gamma_n > 0$) for the energy density of the EM field be 
positive definite (see \cite{Novelloal2000}, appendix B), which entails {\bf B}$ 
> B_{1}$. On the other hand, the good accordance of the Voyager
1/2 magnetometers data with Parker's theory constraints $B_{\rm LSC}$ to be less 
than $0.022~\mu$G within the solar system up to the heliopause. 
Hence, we may conclude that $0.01~\mu$G $< B_{\rm LSC} < 0.022~\mu$G within the 
solar system. By recalling that the uplink frequency of 
Pioneer 10/11 spacecrafts is $\nu = 2.2$~GHz, one obtains for the median value 
$B_{\rm LSC} = 0.018~\mu$G (both expressions are normalized by $\frac{H_0} 
{70~\rm{km} ~{s^{-1} ~Mpc^{-1}}}$, Eq.(\ref{16}))
}

\begin{eqnarray}
\frac{\dot{\nu}}{\nu} = 2.8 \times 10^{-18} ~ {\rm s^{-1}}\, , \hskip 0.5 truecm 
\frac{d\Delta \nu}{dt} = 6 \times 10^{-9} \, \rm \frac{Hz}{s}\, ,
\label{dot-freq}
\end{eqnarray}

with $\Delta \nu$ the frequency discrepancy pointed out earlier.

%% between the modelization taking into account NLED and that based on Maxwell theory.

%%% $B_{\rm LSC} = 1.77~B_1$ .  the CMB is well described by Maxwell
%%% theory which is likely to give a good account of the magnetic fields in
%%% galaxies too. However, the processes at the origin of both the seed magnetic
%%% field in the interclusters medium and clusters of galaxies are not yet
%%% clearly understood. Hence, if NLED is to play a significant role at the
%%% macroscopic scale, this should occur at the intermediary scales of clusters
%%% of galaxies or the interclusters medium. In 

%% Considering the possibility that the NLED correction terms described above come 
%%% into play at these scales, one gets the following ordering of magnitudes : 
%%%$B_{\rm Universe} \ll B_{\rm Intercluster} \ll B_1 \lesssim B_{\rm Cluster} \lesssim B_{\rm Galaxy}$. 
%%% Turning back to the Pioneer anomaly, a good accordance is obtained with $B_{\rm 
%% LSC} = 1.77~B_1$ (case $n=1$){\bf \cite{estimating-gamma}}. 

{A Note on interplanetary magnetic field and NLED effects}.--- 
{ It has been pointed out that the strength of the IPMF could severly 
minimize the NLED effects, { because it will overrun the interstellar 
or intergalactic magnetic fields at heliocentric distances}. Notwithstanding, 
the actual data from Voyager 1/2 spacecrafts of the IPMF average strength 
are both consistent with a non-zero local supercluster magnetic field (LSCMF) 
amounting up to $0.022~\mu$G \cite{voyager1,voyager2} (the accuracy 
of the measurements performed by Pioneer 10/11 magnetometers is at best $0.15~ 
\mu$G, and $0.022~\mu$G for the low field system of Voyager 1/2 magnetometers 
\cite{solar-wind}). Besides, it is just beyond the Saturn orbit, $\sim 10$ 
Astronomical Units (AU), that the anomaly begins to be clearly observed. {  
Surprisingly, it is just after passing the Saturn orbit that the strength
of the magnetic field vehiculated by the solar wind gets down the strength 
associated to the insterstellar and intergalactic magnetic fields, as one can 
verify by perusing Refs.\cite{voyager1,voyager2}. Thus, since a magnetic field 
cannot shield (or block) any another magnetic field (the stronger field can only
reroute the weaker  field, otherwise it would violate Maxwell's laws), then 
it follows that the LSCMF has its magnetic influence extended upto nearly the 
location of the Saturn orbit, and in this way it forces the photons being emitted
by the Pioneer spacecrafts from larger heliocentric distances to get accelerated 
due to the NLED effects.  Besides, notice that Ref.\cite{mitra2003} also shows 
that the local cloud of interstellar gas in HII regions does not keep out the 
Galactic magnetic field.

In passing, we call to the reader's attention the fact that some workers in 
the field have claimed that the effect should 
have showed up already at the small distance corresponding to Mars, Jupiter or 
Saturn orbits, because of the high technology involved in the tracking of planet 
orbiting spacecrafts as Galileo and Cassini or the Mars' nonroving landers, 
which would allow to single out the anomaly at those heliocentric distances. 
However, as those spacecrafts are inside the region where the solar wind dominates, 
this definitely precludes the NLED photon acceleration effect to show up at
those distances since the much higher magnetic field there would introduce 
a negligible NLED effect, and as stated below the solar pressure influence on
the signal frequency is still large. 

%%On the other hand, although one 
%%can use the time of flight of photons during tracking of planets with orbiting 
%%spacecrafts (by combining range and Doppler data over a spacecraft orbit) to 
%%tightly determine the range from Earth to that given planet's center of mass, 
%%the impediment to single out the radio-signal frequency shift remains the same 
%pictured above: from one side the strength of both the host planet magnetic 
%%field and the solar magnetic field at those 
%%distances are  still large, what blocks the action of the LSCMF, and from the 
%%other, the frequency shift due to the solar pressure is still significantly 
%%large up to 20 AU so as to allow the show up of the NLED frequency shift which 
%is much smaller. } 

%5 Moreover, within a heliocentric distance $\sim$ 100~AU the IPMF keeps stronger %% than $0.2~\mu$G (see plots in Refs. \cite{voyager1,voyager2}). This in turn 
%%reduces for all practical purposes the IPMF contribution to the effects 
%%of NLED (see further arguments from our direct estimate of $B_1$ 
%%in Ref.\cite{estimating-gamma}), leaving room for the sole contribution 
%%of the residual IGMF in the solar system.

%%%%%%%%%%%%%%%%%%%%%%%%%%%%%%%%%%%%%%%%%%%%%%%%%%%%%%%%%%%%%%%%%%%%%%%%%%%%%%%%%

Finally, the new frequency shift that is predicted by NLED is not seen yet in 
the laboratory { because of the following reasons: a) the most important, the 
strength of the Earth magnetic field is much larger than the one required in 
the NLED explanation of the anomaly for the effect to show up, and b) } the 
coherence time $\tau = 1/\Delta\nu$ of EM waves in 
present atomic clocks (frequency width $\Delta\nu > 0.01$ Hz, or otherwise 
stated $c\tau < 0.2$ AU) is too short as compared to the time of flight of 
photons from Pioneer 10/11 spacecrafts past 20 AU. Nonetheless, if the 
conditions demanded by our model were satisfied this effect will certainly 
be disentangled in a dedicated experiment { where, for instance, the Earth 
magnetic field is kept outside the case containing an experimental set up
where a very weak magnetic field is maintained inside, a source of photons set 
to travel and a receiver  data-collecting.} 

%and also too noisy (irregularities on the planet surface is one such source of noise that needs a sufficiently big distance to be minimized), also 

}

%%%%%%%%%%%%%%%%%%%%%%%%%%%%%%%%%%%%%%%%%%%%%%%%%%%%%%%%%%%%%%%%%%%%%%%%%%%%

\acknowledgements{The authors thank Dr. Santiago E. P\'erez Bergliaffa and Jacques 
Vall\'ee for fruitful discussions on the matters of this paper, and the support 
of ICRA-BR/CBPF, CNPq and FAPERJ (Brazil).}


\begin{thebibliography}{99}

\bibitem{1} J.  D.  Anderson {\sl et al.}, Phys. Rev. Lett. 81, 2858
(1998)

\bibitem{2}C.  B.  Markwardt, preprint gr-qc/0208046.

\bibitem{3} J.  D.  Anderson {\sl et al.}, Phys. Rev. D 65, 082004 (2002)

\bibitem{8} J.  P.  Mbelek \and M.  Michalski, Int. J. Mod. Phys. D 13, 865 
(2004)

\bibitem{proposals} The Pioneer Explorer Collaboration, report gr-qc/0506139

\bibitem{4} H. B Quevedo, report gr-qc/0501006


%%%%  \bibitem{5}M.  Milgrom, 1983, ApJ 270, 365.

\bibitem{6} M.  Milgrom, Acta Phys. Polon. B32, 3613 (2001), 
New Astron. Rev. 46, 741 (2002)

\bibitem{7}A.  Abramovici \and Z.  Vager, Phys. Rev. D 34, 3240 (1986)

\bibitem{10} D.  P.  Whitmire \and J.  J.  Matese, Icarus 165, 219 (2003)

%%%% \bibitem{9} R.  Foot and R.  R.  Volkas, Phys.  Lett.  B 517, 13 (2001)

%\bibitem{11}B.  G.  Marsden and G.  V.  Williams, 2001, Catalog of
%Cometary Orbits, 14th edition, Smithsonian Astrophysical Observatory,
%Cambridge, MA.

\bibitem{12} A.  F.  Ra\~nada, Found. Phys. 34, 1955 (2005),
Europhys. Lett. 63, 653 (2003)

\bibitem{13} J.  P.  Mbelek, Int. J. Mod. Phys. A 20, 2304 (2005)

{ \bibitem{iorio2006}L. Iorio, report gr-qc/0601055 (2006), report 
gr-qc/0608068 (2006), report gr-qc/0608101 (2006)}

{ \bibitem{tangen2006}K. Tangen, gr-qc/0602089: Could the Pioneer 
anomaly have a gravitational origin? (2006) }

\bibitem{DM}J. D. Anderson, et al., Astrophys. J. 448, 885 (1995)

%%%% \bibitem{riess2004}A. G. Riess, et al., Astrophys. J. 607, 665 (2004)

\bibitem{Novello-Salim2002} M. Novello \and  J. M. Salim, Phys. Rev.
D 63, 083511 (2001)

\bibitem{Novelloal2000} M. Novello, V. A. De Lorenci, J. M. Salim \and
R. Klippert, Phys. Rev. D 61, 045001 (2000)

{ \bibitem{Fujii} Y. Fujii and M. Sasaki, report astro-ph/0608508 (2006) }

\bibitem{plebanski} J. Plebanski. Lectures on nonlinear electrodynamics,
Nordita, Copenhagen, (1970).

\bibitem{NLED-REFRENCES}D. L. Burke, et al., Phys. Rev. Lett. 79, 1626 (1997). 
See also E. Lundstrom, et al., Phys. Rev. Lett. 96: 083602 (2006); 
J. Lundin et al., e-Print Archive: hep-ph/0606136; M. Novello et al. Phys. Rev. 
D 61:045001 (2000); V. A. De Lorenci et al., Phys. Lett. B 482: 134-140 (2000);
H. J. Mosquera Cuesta, J. A. de Freitas Pacheco and J. M. Salim Int. J. Mod. Phys. 
A 21: 43-55 (2006), J. T. Mendonca, et al., e-Print Archive: hep-ph/0607195; and 
the complete review by M. Marklund, P. K. Shukla, Rev. Mod. Phys. 78: 591-640 (2006)

\bibitem{hadamard}Following Hadamard \cite{HAD}, the surface of 
discontinuity of the EM field is denoted by $\Sigma$. The field 
is continuous when crossing $\Sigma$, while its first derivative 
presents a finite discontinuity. These properties are specified 
as follows: $\left[F_{\mu\nu} \right]_{\Sigma} =  0\, ,$ \hskip 
0.3 truecm $\left[F_{\mu\nu|\lambda}\right]_{\Sigma} = f_{\mu\nu}
k_\lambda \, \protect \label{eq14} \,$, where the symbol
$\left[F_{\mu\nu}\right]_{\Sigma} = \lim_{\delta \to 0^+}
\left(J|_{\Sigma + \delta}-J|_{\Sigma - \delta}\right)$
represents the discontinuity of the arbitrary function $J$
through the surface $\Sigma$. The tensor $f_{\mu\nu}$ is
called the discontinuity of the field,  $k_{\lambda} =
\partial_{\lambda} \Sigma $ is the propagation vector, and
symbols "$_|$" and "$_{||}$" stand for partial and covariant 
derivatives.

\bibitem{HAD} Hadamard, J., Le\c cons sur la propagation des ondes
et les \'equations de l'Hydrodynamique (Hermann, Paris, 1903)

\bibitem{GALACTIC-B-FIELDS}The value of the magnetic field measured in
the Galaxy is $B_{\rm Galaxy} = 6 \pm 2 \times 10^{-6}$~G \cite{beck2000} 
and that of the local cluster of galaxies is $B_{\rm Cluster} \sim  2-3 
\times 10^{-6}$~G \cite{beck2000}, and for extragalactic clusters as Abell 
400, Hydra A measurements provide $B_{\rm EG-C} \sim  3-20 \times 10^{-6}$~G 
\cite{vogt-ensslin2003}.

\bibitem{vogt-ensslin2003} C. Vogt \and T. A. Ensslin, Astron.Astrophys. 412, 
373 (2003). See also K. Dolag, M. Bartelmann and H. Lesch, Astron. Astrophys. 
348, 351 (1999)

\bibitem{beck2000} R. Beck, Astrophys. Sp. Sc. Rev. 99, 243, published in
Proceedings of {\sl The Astrophysics of Galactic Cosmic Rays} Conference,
eds. R. Diehl et al., Kluwer, Dordrecht (2001)

\bibitem{Vallee} J. P. Vall\'ee, Astronom. J., 124, 1322 (2002)

\bibitem{BB1970}Z. Bialynicka-Birula \and I. Bialynicki-Birula,
Phys. Rev. D 2, 2341 (1970)

\bibitem{approximation}Given a background metric
$g_{\mu\nu}$, as a result of NLED effects photons follow geodesic 
paths with respect to the effective metric (or any one conformal to it)
$g^{(eff)}_{\mu\nu} = g_{\mu\nu} - 4\frac{L_{FF} }{L_{F} } F_{\mu}^{\, \, 
\alpha} F_{\alpha\nu}$ (see \cite{Novelloal2000},\cite{Novello-Salim2002}).
Thus, following the usual analysis on the gravitational frequency shift 
but with $g^{(eff)}_{\mu\nu}$ replacing $g_{\mu\nu}$, one gets $k^{0} c = 
{\omega}_{0}/\sqrt{g^{(eff)}_{00}}$ (see \cite{LL1970}, section 88), 
where ${\nu}_{0} = {\omega}_{0}/2\pi$ denotes the photon frequency in 
flat Minkowski spacetime. Thus, discarding the cosmological redshift 
(subsequent to the time dependence of the curvature), the variation of 
$k^{0}$ with time can be neglected in the first order approximation, 
since $F^{0\alpha} F_{\alpha}^{\, \, 0} =  F_{0}^{\, \, \alpha}
F_{\alpha 0} = 0$ in the case of a zero electric field.

\bibitem{NSB2004} M. Novello \and S. E. P\'erez Bergliaffa \and J. M. Salim,
Phys. Rev. D 69, 127301 (2004).

\bibitem{LL1970} L. D. Landau \and E. Lifchiftz, Th\'eorie des Champs,
Editions MIR, Moscou (1970)

%%%\bibitem{NB2005} 

{ 
\bibitem{estimating-gamma}A clean estimate of $B_1$ from our definition 
of $\gamma_n \equiv - (B_n c)^{2(n+1)}$, below Eq.(\ref{14}), and the one 
in Ref.\cite{NSB2004}: $\gamma \equiv - \hbar^{2} ~\mu^{8}$. On account 
that $a_{c} = (1 + z_{c})^{-1}$ and $\gamma = -  \hbar^{2} ~\mu^{8} = - (B_{1} 
c)^{4}$, Eq.(13) of Ref.\cite{NSB2004} rewrites $B_{1}^{4}/\mu_{0}~B_{0}^{2} 
= 1.40 \rho_{c} c^{2}$ (A1). Thus $a_{c} = (B_{0}^{4}c^{4}/3\hbar^{2}\mu^{8})^{1/8}$ 
yields $B_{1} = 3^{-1/4} (1 + z_{c})^{2} B_{0}$ (A2). Then, combining both 
relations (A1) and (A2) one gets : $B_{0} = 0.02~(1 + z_{c})^{-4}$~$\mu$G (A3),
$B_{1} = 0.016~(1 + z_{c})^{-2}$~$\mu$G (A4). Now, Riess et al. found evidence 
for a transition from cosmic deceleration to acceleration at redshift $z_{c} = 
0.46 ~\pm ~0.13$ [A. G. Riess et al., ApJ 607, 665 (2004)]. Inserting the latter 
figures in relations (A3) and (A4)  yields : $B_{0} = (0.005 ~\pm ~0.002)$~
$\mu$G (A5), $B_{1} = (0.008 ~\pm ~0.002)$~$\mu$G (A6). Since CMB is pure 
radiation (i. e., $E = B c$ not equal to zero on average), we consider that 
relations (A4) and (A6) give a better estimate of $B_{1}$ than the one put 
forward in Ref.\cite{NSB2004}.
}

{
\bibitem{Blasi1999} P. Blasi \and A. V. Olinto, Phys. Rev. D 59, 023001 (1999)
}
 
{ 

\bibitem{voyager1} Voyager 1 Observations and Parker's Model 
of Interplanetary Magnetic Field (IPMF) Strength, at: 
http://spacephysics.ucr.edu/images/swq2$_{-}$04.jpg , 
http://www.phy6.org/Education/wtermin.html

\bibitem{voyager2}Voyager 2 Observations and Model Prediction of IPMF Strength, 
at: http://interstellar.jpl.nasa.gov/interstellar/probe/interstellar/images/07BupstrmSuess.25$_{-}$lg.gif

%%%% {mathewsson68} Mathewsson and Nicholls, 1968, Astrophys. J. 154, L11

\bibitem{solar-wind} Solar Wind: Detection Methods and Long-term Fluctuations, 
at: http://herkules.oulu.fi/isbn9514271955/isbn9514271955.pdf

\bibitem{mitra2003}D. Mitra, et al., Astron. Astrophys. 398, 993 (2003)

}

%%% R. N. Manchester, 1973, Astrophys. J 172, 43

\end{thebibliography}
\end{document}